\documentclass{jfm}

\usepackage{graphicx}
\usepackage{bm}
\usepackage{amsmath}
\usepackage{cleveref}
\usepackage[OliveGreen,dvipsnames]{xcolor}
\usepackage[normalem]{ulem}

\newcommand{\mbq}{\overline{{\bf q}}}
\newcommand{\dmbq}{\delta \overline{{\bf q}}}

\newcommand{\fq}{q^\prime}
\newcommand{\fbq}{{\bf q}^\prime}

\newcommand{\dfbq}{\delta {\bf q}^\prime}

\newcommand{\opL}{{\cal L}}
\newcommand{\J}{{\cal J}}
\newcommand{\D}{{\cal D}}
\newcommand{\N}{{\cal N}}

\newcommand{\cz}{c_{\zeta}}
\newcommand{\cs}{c_{\psi}}

\newcommand{\css}{c_{\psi \psi}}
\newcommand{\csz}{c_{\psi \zeta}}
\newcommand{\czs}{c_{\zeta \psi}}
\newcommand{\czz}{c_{\zeta \zeta}}

\def\NEW#1{\textcolor{black}{#1}}

\shorttitle{Non-equivalence of QL and CE2}
\shortauthor{Nivarti, Kerswell, Marston and Tobias}

\title{Non-equivalence of quasilinear dynamical systems and their statistical closures}
\author{G. V. Nivarti\aff{1}\corresp{gvn22@cantab.ac.uk} \and R. R. Kerswell\aff{2} \and J. B. Marston\aff{3} \and S. M. Tobias\aff{1}}

\affiliation{\aff{1}Department of Applied Mathematics, University of Leeds, UK \aff{2}Department of Applied Mathematics and Theoretical Physics, University of Cambridge, UK \aff{3}Brown Theoretical Physics Center and Department of Physics, Brown University, Providence, RI 02912-S USA}

\begin{document}

\maketitle
\begin{abstract}
It is widely believed that statistical closure theories for dynamical systems provide statistics equivalent to those of the governing dynamical equations from which the former are derived. Here, we demonstrate counterexamples in the context of the widely used mean-field quasilinear (QL) approximation applied to $2$D fluid dynamical systems. We compare statistics of QL numerical simulations  with those obtained by direct statistical simulation via a cumulant expansion closed at second order (CE2). We observe that, though CE2 is an exact statistical closure for QL dynamics, its predictions disagree with the statistics of the QL solution for identical parameter values. These disagreements are attributed to instabilities, \NEW{which we term rank instabilities}, of the second cumulant dynamics within CE2 that are unavailable in the QL equations.
\end{abstract}


\maketitle

\section{Introduction}
A common problem in fluid dynamics, and indeed nonlinear physics in general, is to obtain statistical descriptions for quantities whose master dynamical equations are known. One approach is to accumulate statistics from numerical solutions of the master equations; for example one may solve the Navier-Stokes equations for a turbulent flow and gather statistics via, for example, time averaging, ensemble averaging or averaging over a spatial co-ordinate. An alternative approach, termed Direct Statistical Simulation (DSS) seeks solutions of closed-form equations for the statistics themselves that are derived from the dynamical equations; these may be equations that describe the probability density function (pdf) of the system or the low-order statistics thereof \citep[][]{Allawala2016}. In many contexts, particularly in fluid dynamics, these statistical approaches are more favourable computationally; they can alleviate the stringent requirements of spatial resolution and sample sizes. The underlying expectation in this approach, however, is that solutions of the statistical equations will match those obtained from solutions of the dynamical equation.

Here we consider the specific case of quasilinear (QL) theory, a type of mean-field approximation that has been widely applied to a range of physical systems in fluids (and indeed plasmas) \citep{Malkus:1954dh,Vedenov:1961us,Herring:1963}. In recent years, QL equations have been used to simulate zonal jets -- strong east-west mean flows -- found in many astrophysical and geophysical situations \citep{Farrell2007, Young12, Galperin2019} as well as to simulate flows in rotating dynamos \citep{cjta2015}, plasmas \citep{parkerkrommes2014}, wall-bounded shear flows \citep{Farrell16} and convection \citep{OConnor21}. The simplest non-trivial statistical closure -- called CE2 -- naturally follows the assumption of quasilinearity by retaining only up to second order cumulants, and has enabled efficient DSS \citep{Marston2008,Marston2010,Tobias2011,Marston2019}. Also see \cite{Farrell2007,Farrell2013} for the closely related closure Stochastic Structural Stability Theory (SSST) where stochastic forcing replaces the neglected cumulants.

It is often supposed (usually implicitly) that the solution of the statistical equations should mirror those of the dynamical system. In this paper, we demonstrate that this is not always the case for quasilinear dynamics in a range of fluid problems, solved via numerical computation. We associate the discrepancy between the statistical theory (CE2) and a given realisation of the QL dynamics to a ``rank instability" of the CE2 system that is not available to QL.

This paper is organised as follows. In \S2 the fluid models used as a testbed are described and the numerical framework is set up. In \S3 results are presented for deterministic and stochastically forced systems, showing the potential discrepancy between QL and CE2. Theory for the origin of the discrepancy including an exactly solvable illustration is given in \S4 and conclusions are drawn in \S5.

\section{Set-up of the Models and Numerical framework}

\subsection{The fully nonlinear model}

We conduct two-dimensional simulations of a rotating, incompressible fluid with velocity ${\boldsymbol u}$ on a doubly-periodic $\beta$-plane $[0,2\pi]^2$. The time evolution of the relative vorticity $\zeta \equiv \widehat{z}\cdot (\nabla \times {\boldsymbol u})$ is given by
\begin{equation}
    \label{eq:vorticity}
    \partial_t \zeta = \beta \partial_x \psi - \kappa\zeta + \nu \nabla^2 \zeta + J[\psi,\zeta] + F,
\end{equation}
where the Jacobian $J[\psi,\zeta] \equiv \partial_x\psi \partial_y \zeta - \partial_x \zeta \partial_y \psi$ and the streamfunction $\psi$ is given by $\psi \equiv \nabla^{-2}\zeta$. Gradients of rotation are included via the $\beta$ term, `bottom' friction via the $\kappa$ term and viscosity via the $\nu$ term. We adopt \NEW{three different} forcing models to enable the comparison of QL and CE2 with different dynamics. The first two forcings considered are deterministic. In the first, the flow is made to relax to an unstable jet profile \citep{Marston2008}: see \S\ref{F1}. In the second a steady, two-scale Kolmogorov-type forcing is used \citep{Tobias2017}: see \S\ref{F2}. \NEW{The third forcing we consider is stochastic in nature \citep{Constantinou2016}: see \S\ref{F3}.}

\subsection{The quasilinear model}
The quasilinear model of the fully nonlinear system is obtained by decomposing the variables into a mean and fluctuating part, i.e. we write
\begin{equation}
\zeta = \overline{\zeta}+ \zeta^\prime; \quad \psi = \overline{\psi} + \psi^\prime,
\end{equation}
where overbar indicates a zonal average given by
\begin{equation}
\overline{f}(y)= \frac{1}{2 \pi}\int_0^{2 \pi} f(x,y) \, dx; \quad f^\prime(x,y) = f(x,y) - \overline{f}(y).
\end{equation}
It is then possible to derive the quasilinear model by considering the interactions between the means and the fluctuations so that
\begin{equation}
    \label{eq:vorticity_mean}
    \partial_t \overline{\zeta} = \beta \partial_x \overline{\psi} - \kappa \overline{\zeta} + \nu \partial_{yy} \overline{\zeta} + \overline{J[\psi^\prime,\zeta^\prime]} + \overline{F},
\end{equation}
and
\begin{equation}
    \label{eq:vorticity_fluc}
    \partial_t \zeta^\prime = \beta \partial_x \psi^\prime - \kappa\zeta^\prime + \nu \nabla^2 \zeta^\prime + J[\overline{\psi},\zeta^\prime] + J[{\psi^\prime},\overline{\zeta}] + \left(J[\psi^\prime,\zeta^\prime] - \overline{J[\psi^\prime,\zeta^\prime]}\right)+ F^\prime,
\end{equation}
and \NEW{then} removing the \NEW{fluctuation-fluctuation (or `eddy-eddy')}  interactions (in \NEW{rounded} brackets) to yield the quasilinear system
\begin{equation}
    \label{eq:vorticity_mean_ql}
    \partial_t \overline{\zeta} = \beta \partial_x \overline{\psi} - \kappa \overline{\zeta} + \nu \partial_{yy} \overline{\zeta} + \overline{J[\psi^\prime,\zeta^\prime]} + \overline{F},
\end{equation}
\begin{equation}
    \label{eq:vorticity_ql}
    \partial_t \zeta^\prime = \beta \partial_x \psi^\prime - \kappa\zeta^\prime + \nu \nabla^2 \zeta^\prime + J[\overline{\psi},\zeta^\prime] + J[{\psi^\prime},\overline{\zeta}] + F^\prime.
\end{equation}
It is this system that will be compared with the cumulant expansion (CE2) described below.

\subsection{The cumulant expansion at second order (CE2)}

The CE2 system is derived by first defining the cumulants. The first cumulant for the vorticity is given by $\cz(y) = \overline{\zeta}$,
%
%
whilst the second cumulants are defined by
\begin{equation}
    c_{\zeta\zeta}(\xi,y_1,y_2) = \overline{\zeta^\prime(x_1,y_1) \zeta^\prime(x_1-\xi,y_2)}
\end{equation}
where $\xi = x_1 - x_2$ \NEW{as the system is translationally invariant in $x$}. Here $c_{\zeta \zeta}$ can be directly related to $c_{\zeta \psi}$ and $\csz$ by differentiation; for example $\czs = \nabla^2_1 \css$ where the differential operators $\nabla^2_1 = \partial_\xi^2 + \partial_{y_1}^2$ and $\nabla^2_2 = \partial_\xi^2 + \partial_{y_2}^2$ \citep[see][for more detail]{Tobias2011}. We can then proceed by
writing the  equations in terms of $\zeta$ cumulants as follows.

The  dynamical equations for the first cumulant $\cz$ is given by
\begin{equation}
  \label{eq:cz}
  \dfrac{\partial \cz}{\partial t} = -\kappa \cz +\nu \dfrac{\partial^2 \cz}{\partial y_1^2} - \left(\dfrac{\partial}{\partial y_1} + \dfrac{\partial}{\partial y_2}\right) \dfrac{\partial \csz}{\partial \xi}\Big|_{\xi \to 0}^{y_1 \to y_2} + \NEW{\overline{F}}
\end{equation}
whilst that for the second cumulant $\czz$ is given by
\begin{equation}
  \label{eq:czz}
  \begin{split}
    \dfrac{\partial \czz}{\partial t} &= \dfrac{\partial \cs}{\partial y_1} \dfrac{\partial \czz}{\partial \xi} - \left(\dfrac{\partial \cz}{\partial y_1}-\beta \right) \dfrac{\partial \csz}{\partial \xi} - \dfrac{\partial \cs}{\partial y_2} \dfrac{\partial \czz}{\partial \xi}  + \left(\dfrac{\partial \cz}{\partial y_2}-\beta \right)  \dfrac{\partial \czs}{\partial \xi}\\
   &  \hspace{6cm} +\nu (\nabla_1^2 + \nabla_2^2) \czz - \kappa \czz + \Gamma,
  \end{split}
\end{equation}
Here $\Gamma$ is given by the covariance of the \NEW{stochastic} forcing\NEW{, i.e. $\Gamma(\xi, y_1, y_2) = \overline{F^\prime ({x_1},y_1)F^\prime (x_1-\xi,y_2)}$} \NEW{with the zonal average over $x_1$ combined with a short time average over the fast stochastic fluctuations.} \NEW{The QL system defined by \cref{eq:vorticity_mean} and \cref{eq:vorticity_fluc} can be transformed into wavevector space in a basis of Fourier harmonics. Similarly, the CE2 system defined by \cref{eq:cz} and \cref{eq:czz} above can be solved \NEW{in} Fourier space by time-evolving $\hat{c}_\zeta(n)$ and $\hat{c}_{\zeta \zeta} (m, n_1, n_2)$ for the first and second cumulants, respectively. Here $m$ is the wavenumber in the zonal ($x$) direction and $n$, $n_1$ and $n_2$ are wavenumbers in the non-zonal ($y$) direction.  For convenience in the following we define the matrix $C^{(m)}$ to be the zonal decomposition of the second cumulant: $C^{(m)} = \hat{c}_{\zeta \zeta} (m, n_1, n_2) = \frac{1}{(2\pi)^3}\int_{0}^{2\pi}\int_{0}^{2\pi}\int_{0}^{2\pi}c_{\zeta\zeta}(\xi,y_1,y_2) e^{-i m\xi - i n_1y_1 - in_2y_2}d\xi dy_1dy_2$ for $m \in [1,M]$ where $M$ is the spectral cutoff in the zonal wavenumber .}

\subsection{Forcing and Numerical Scheme}

We first examine cases that demonstrate  divergences between QL and CE2 in deterministic systems, before adopting a stochastic forcing model \citep{Young12,Farrell2013,Tobias2013,Constantinou2016,Marston2019}. This enables us to  highlight that  disagreements  between QL and CE2 may also appear in this commonly utilised scenario.

The spectral solver \verb+ZonalFlow.jl+ written in the Julia programming language \citep{Julia2017} and available online \citep{Nivarti2021} is used to obtain QL and CE2 solutions of equation (\ref{eq:vorticity}). Timestepping algorithms are imported from the well-tested ecosystem of the \verb+DifferentialEquations.jl+
package \citep{Rackauckas2017}; here we use the explicit $5/4$ Runge-Kutta method of Dormand-Prince with a fixed timestep for the deterministic cases, and a SRIW1 method of order $1.5$ for the stochastic cases. For the purpose of comparing QL and CE2 solutions, the key is to use the same spatial resolution for each model. \NEW{We use small numbers} of spectral modes -- here $M \times N = 8 \times 8$, $16 \times 16$ and $12\times20$ -- for the Kolmogorov flow, pointjet and stochastic cases respectively. This allows us to demonstrate clearly the differences between QL and CE2.  \NEW{We have verified that} the solution behaviour remains qualitatively similar for higher resolutions.

\section{Results}

\subsection{Deterministic Forcing}

\subsubsection{Relaxation to a pointjet}
\label{F1}
We begin with the case of relaxation to a pointjet. Here the forcing is chosen to be
$F(y) = - ({\Xi}/{\tau}) \rm{tanh}[({\pi - y})/{\Delta y}]$. The $\beta$ plane is equatorial, that is  $\beta = 2\Omega \rm{cos}(0^\circ) = 4\pi$ where the period of rotation $\Omega = 2 \pi$. Following \cite{Marston2008} and \cite{Allawala2020}, the viscosity $\nu = 0$, and the relaxation timescale $\tau$ is set equal to the friction timescale $\kappa^{-1}$. The jet strength $\Xi = \Omega = 2 \pi$ and jet width $\Delta y = 0.1$. For the results shown, $\tau = 20$ days though we stress that the comparisons hold for a wider range of  $\tau$.

\begin{figure*}
    \includegraphics[width=\textwidth]{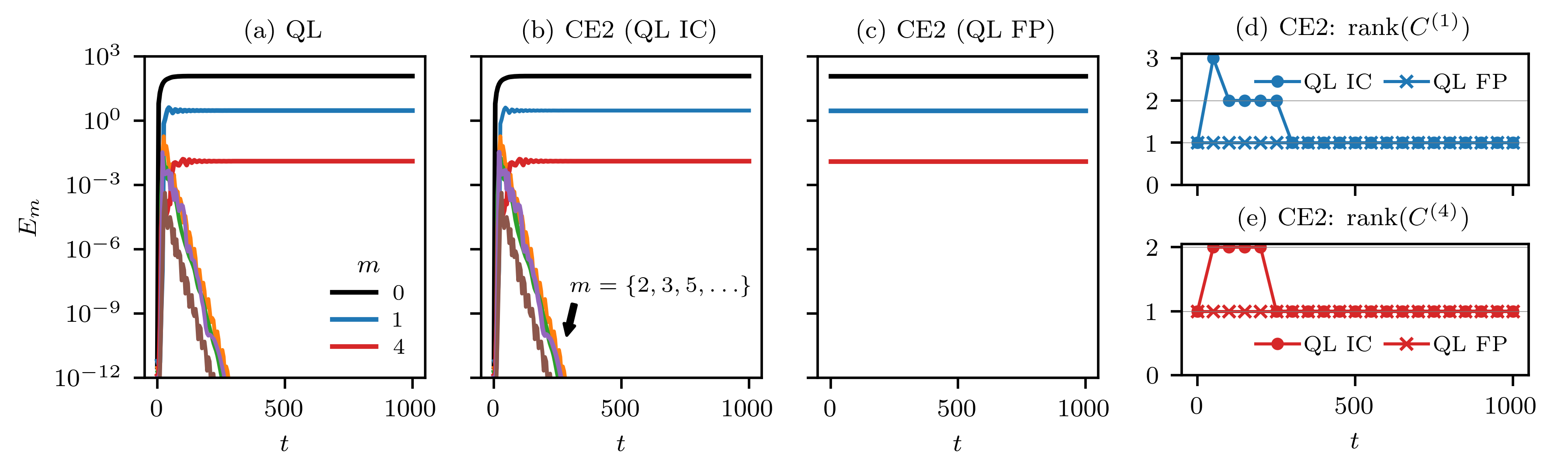}
    \caption{Left: energy in zonal wavenumbers $E_m$ for unit-rank initialisation in the pointjet case up to a spin-up of $1000$ days. Results are for (a) QL, (b) CE2 {initialised with the QL IC, and (c) CE2 initialised with the QL fixed point solution (QL FP).} Right: rank of cumulant submatrices (d) $C^{(1)}$ and (e) $C^{(4)}$ {as found in the two CE2 solutions}. }
    \label{fig:pj:qlce2}
\end{figure*}

\Cref{fig:pj:qlce2} shows both QL and CE2 solutions obtained for this case.
Panels {\it a} through {\it c} show the time evolution of the energy in zonal modes $m$, given by \NEW{ $E_m = \sum_n \hat{\zeta}_{m,n}^*\hat{\zeta}_{m,n}/(m^2 + n^2)$ where $\hat{\zeta}_{m,n}$ is the relevant Fourier coefficient of vorticity $\zeta(x,y)$, which can be calculated in QL using $\zeta = \overline{\zeta}+ \zeta^\prime$. In CE2, $E_m$ can be calculated as $E_m = \sum_n |\hat{c}_\zeta(n)|^2/n^2$ for $m = 0$ and $E_m = \sum_n C^{(m)}(n,n) /(m^2 + n^2)$ for $m \ge 1$.} The QL solution, initialised with a random noise {of strength $10^{-6}$}, predicts a fixed point (FP) with significant energy in the zonal mean ($m = 0$) mode and energy in two further non-zero zonal wavenumbers ($m = 1$ and $m = 4$). We construct two different initial conditions for CE2; in the first the  cumulants are constructed using the QL initial condition (\cref{fig:pj:qlce2}b). For the second initial condition we initialise the CE2 evolution using the stable QL fixed point (\cref{fig:pj:qlce2}c) \citep{Marston2019}. Evidently, for this case, CE2 solutions for both initial conditions agree with the QL fixed point as expected.

\begin{figure}
    \centering
    \includegraphics[scale=0.9]{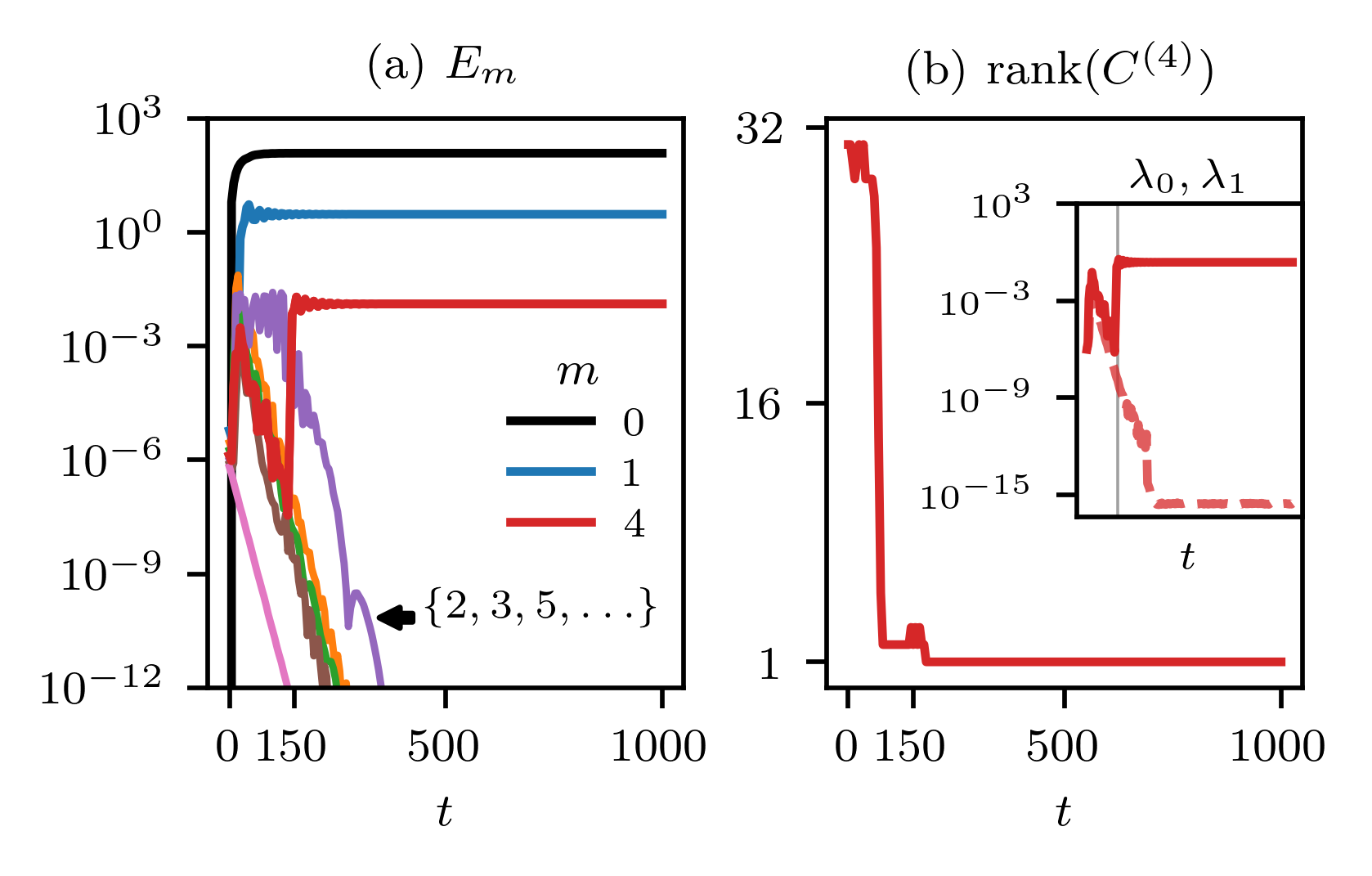}
    \caption{(a) Energy in zonal wavenumbers $E_m$ and (b) evolution of the rank of cumulant submatrix $C^{(4)}$ with full-rank initialisation for the pointjet case {(inset shows evolution of its two largest eigenvalues: $\lambda_0$ and $\lambda_1$). The stability of the unit rank solution is consistent with the agreement of CE2 and QL (\cref{fig:pj:qlce2}a)}.}
    \label{fig:pj:fr}
\end{figure}

Since both initial conditions (ICs) are constructed from a specific QL solution, the cumulant submatrices $C^{(m)}$ have rank unity at initialisation $\forall m \leq M$. Numerically, we calculate this rank as the number of eigenvalues of the second cumulant larger than a small cutoff $10^{-12}$ of the order of the initial power. Notably, the ranks of $C^{(m)}$ for $m = 1,4$ {\it remain unity} for both CE2 solutions $\forall t \geq 0$ as shown in panels (d) and (e), respectively indicating that the statistical solutions are consistent with those of a single realisation of the QL system for the pointjet solution. Thus, this is an example where QL and CE2 are consistent.
%
%
This agreement persists even if the second cumulant is initialized to be full rank in CE2 by assigning a value $10^{-6}$ to diagonal entries in $C^{(m)} \;  \forall m \leq M$. The first cumulant is initialised at zero. \Cref{fig:pj:fr}a shows $E_m$, with the zonal mean ($m = 0$) and first harmonic ($m = 1$) rapidly attaining values close to the QL time mean solution. The second harmonic ($m = 4$) however goes through an initially prolonged period of decay, unlike the unity rank initialisation in \cref{fig:pj:qlce2}b. After about $t = 150$ days, the $m = 4$ mode finally drops to unit rank (\cref{fig:pj:fr}b inset shows evolution of the largest two eigenvalues for $C^{(4)}$). This prompts the QL time mean solution to become stable once again. Thus, despite an initial full rank perturbation, CE2 is able to `find' agreement with QL by virtue of the stability of the unit rank solution; the solution drops rank until its rank is unity and thereafter the agreement between CE2 and QL is ensured. {We stress that all the DSS solutions remain realisable, with positive eigenvalues for the second cumulant.}
%
%
\subsubsection{Kolmogorov Forcing}
\label{F2}
\begin{figure*}
    \includegraphics[width=\textwidth]{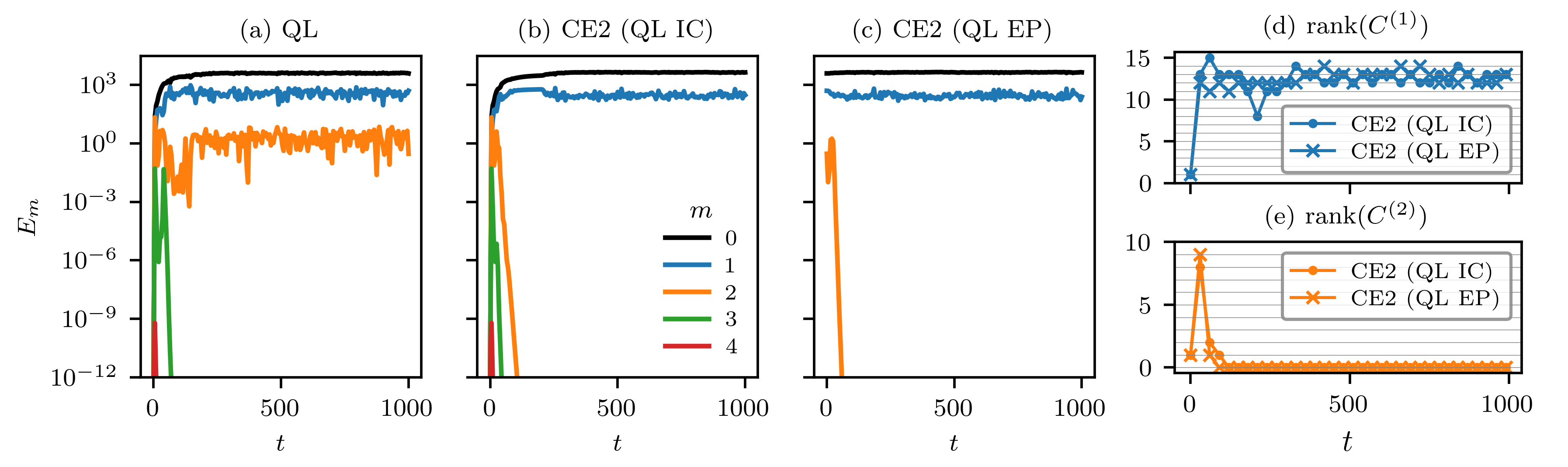}
    \caption{Left: energy in zonal wavenumbers for the Kolmogorov flow case with unity rank initialisation. Results are for (a) QL and (b) CE2 {initialised with the QL initial solution (QL IC), and (c) CE2 initialised with the QL endpoint solution (QL EP).} Right: rank of second cumulant submatrices (d) $C^{(1)}$ and (e) $C^{(2)}$ {as found in the CE2 solutions.}}
    \label{fig:kf:qlce2}
\end{figure*}

Now we give what we believe is the first example of disagreement. \Cref{fig:kf:qlce2} shows solutions for a case with Kolmogorov flow forcing $F(y) = -\cos y -8\cos 4y$, where we choose $\nu = 0.02$ and $\beta = \kappa = 0$ in (\ref{eq:vorticity}), which yields non-trivial dynamics in the resulting system \citep{Tobias2017}. QL (\cref{fig:kf:qlce2}a) predicts a long time solution with non-zero mean and two harmonics, i.e. $m = 1$ (blue) and $m = 2$ (orange). As for the pointjet case, CE2 simulations are performed with unit rank initialisations of the second cumulant using the QL IC (\cref{fig:kf:qlce2}b) and the QL endpoint (EP) solution (\cref{fig:kf:qlce2}c). Here, CE2 finds a single harmonic $m = 1$ in both cases, which disagrees with QL (\cref{fig:kf:qlce2}e shows the rank of $C^{(2)}$ collapsing to zero). The disagreement is caused by the increase in rank of  $C^{(1)}$ from unity to nearly fully rank as shown in \cref{fig:kf:qlce2}d, indicative of a ``rank instability''.
%
%
%
%
\begin{figure}
    \centering
    \includegraphics{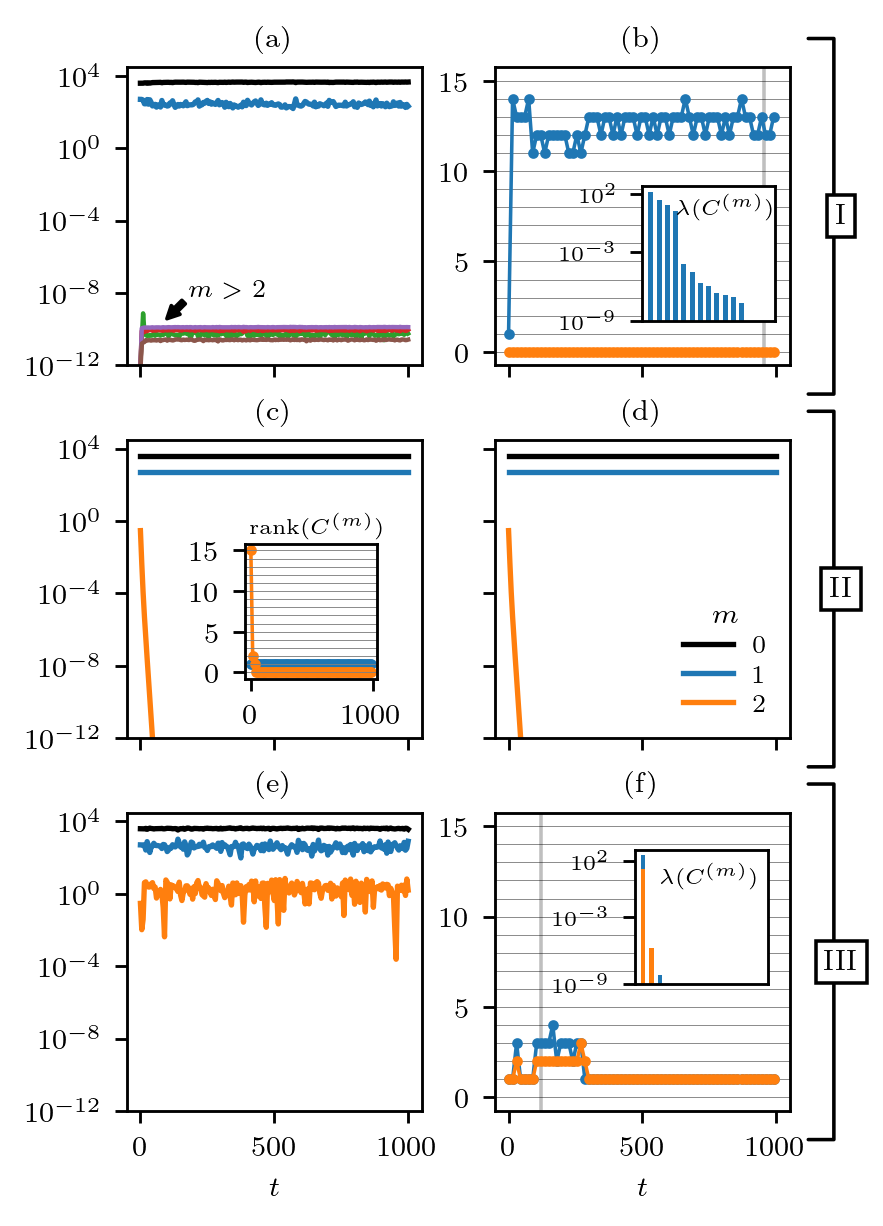}
    \caption{Left column: energy $E_m$ in zonal wavenumbers for CE2 cases I, II and III. Right column, panels b and f: rank of cumulant submatrices $C^{(m)}$ with $m = 1,2$ and panel d: QL solution for case II. The {rank instability} is isolated in case I, which disagrees with the QL EP of \cref{fig:kf:qlce2}a. Case II suppresses the rank instability by locking the base state, resulting in consistent QL and CE2 solutions. Case III recovers the QL EP using eigenvalue truncation (insets in (b) and (f) show spectra at indicated time). }
    \label{fig:kf:ranks}
\end{figure}

To probe this disagreement, we re-run the CE2 simulation initialised using the QL {EP solution} with three separate modifications. First, in case I (\cref{fig:kf:ranks} panels a and b), energy is removed from the zonal wavenumber $m = 2$ in the IC so that the submatrix $C^{(2)}$ has zero rank initially {with the rest of the initial conditions unchanged}. Further, a unit rank random noise (power $10^{-12}$) is added to wavenumbers with $m > 2$. The resulting CE2 solution (\cref{fig:kf:ranks}a)  disagrees with the QL EP (\cref{fig:kf:qlce2}a), the rank instability appearing again in $m =1$: see (\cref{fig:kf:ranks}b). Eigenvalue spectra of $C^{(1)}$ (inset of \cref{fig:kf:ranks}b) confirm that the additional eigenvalues contributing to its higher rank are non-zero. Hence, the disagreement observed {between the CE2 solution} and the QL solution in case I is attributable to a rank instability in  $m=1$.

Second, in case II (\cref{fig:kf:ranks} panels c and d) starting with the QL EP solution, the $m=0$ and $m=1$ components are frozen, thereby ensuring their unit ranks, while the $m = 2$ harmonic is assigned a small full rank perturbation.
As previously, for CE2, the $m=2$ component decays away (\cref{fig:kf:ranks}c). Repeating the same experiment for QL (\cref{fig:kf:ranks}d), also now sees the $m=2$ component decay in agreement with CE2. This indicates that, if the same base state is considered for both systems, the zonal stability characteristics are the same.

Third, in case III (\cref{fig:kf:ranks} panels e and f), the QL EP is again used as the IC, but during time integration only the largest eigenvalue is retained for each second cumulant submatrix $C^{(m)} \; \forall m \leq M$. This ensures that unit rank is maintained at all times for the entire second cumulant (eigenvalue spectra in the inset of \cref{fig:kf:ranks}f confirm that the initially observed departures are trivial). The resulting CE2 solution for case III (\cref{fig:kf:ranks}e) is now in agreement with the QL {solution (\cref{fig:kf:qlce2}a)}. Thus, the CE2 solutions in cases II and III, though exhibiting different zonal stability characteristics, agree with their corresponding QL solutions if the rank of the second cumulant is artificially kept at unity; it is the rank instability in $m=1$ that causes the observed divergence between CE2 and QL. As the rank of the $m=1$ harmonic grows beyond unity, there are more channels available for wave energy to flow depriving the $m=2$ harmonic of any energy. As a check on our simulations, we have reproduced the rank and zonal instabilities on the sphere with independently developed code that uses an \NEW{adaptive RK4 time integration algorithm \citep{pmt2019,MT2023}}.  The Kolmogorov forcing employed is $F(\theta) = -P_2(\cos{\theta}) - 8 P_8(\cos{\theta})$, where $P_\ell$ are Legendre polynomials and $\theta$ is the co-latitude.

\subsection{Stochastic Forcing}
\label{F3}
The above examples indicate that there may be disagreement between the numerical solutions of the QL and CE2 systems. Relying on the determinism of the cases considered thus far, we have linked the disagreement to (1) a rank instability permitted only in CE2, and (2) to a possible subsequent zonal mode instability arising in QL but not in CE2. Our example shows both causes of divergence but the rank instability is the key which may or may not trigger a subsequent difference in the zonal instability characteristics of the two systems.
%
%
\begin{figure*}
    \centering
    \includegraphics[width=\textwidth]{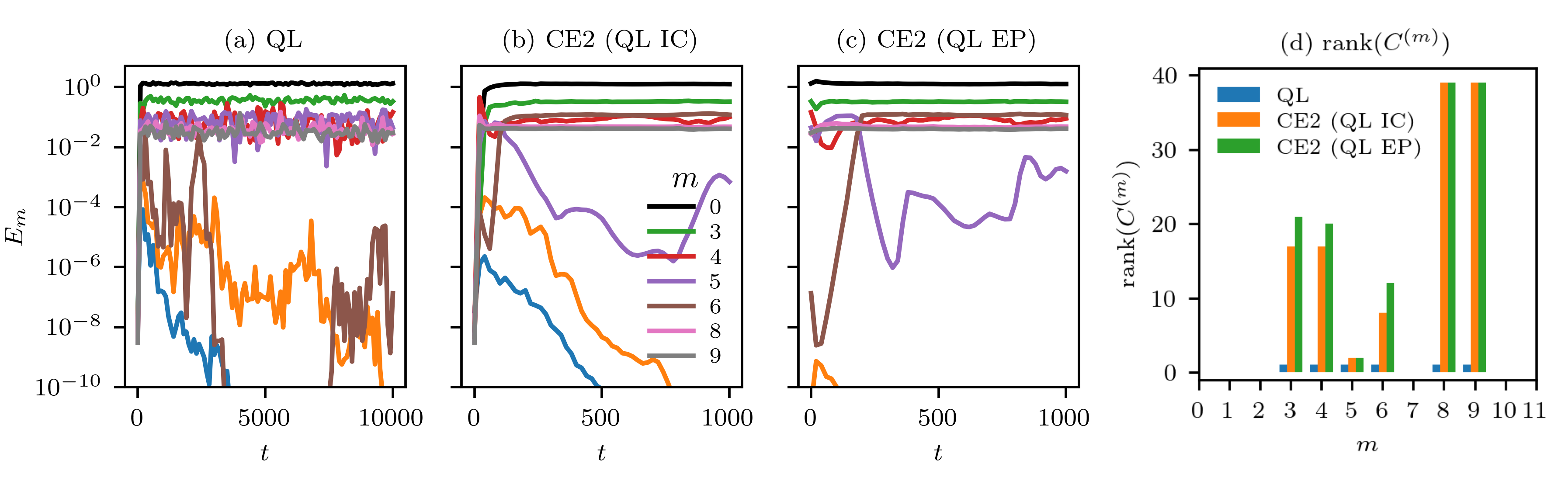}
    \caption{Left: energy in zonal wavenumbers for the stochastically-forced case with unit rank and full rank initialisation. Results are for (a) QL and (b) CE2 initialised {using the QL initial solution (QL IC), and (c) CE2 initialised using the QL end point solution (QL EP).} Right: the corresponding rank of second cumulant submatrices (d).}
    \label{fig:sf:qlce2}
\end{figure*}

We now show that divergences arising from rank instabilities are not limited to deterministic systems. Adopting the stochastic forcing strategy detailed in \cite{Constantinou2015} and \cite{Constantinou2016}, the forcing term in (\ref{eq:vorticity}) becomes \NEW{$F = \sqrt{\varepsilon} ~{\eta}(x,y)$, where $\varepsilon$ is the energy injection rate and ${\eta}$ is the stochastic noise with variance $Q(x,y)$. Hence, $\overline{F} = \overline{{\eta}} = 0$ and the forcing term $F^\prime = F = \sqrt{\varepsilon} ~{\eta}(x,y)$ in \cref{eq:vorticity_mean} drives the QL system. The CE2 system is driven by the forcing term in \cref{eq:czz} which becomes $\Gamma = \varepsilon~ Q ( x, y)~$, where $Q(x, y) = Q(x_1 - x_2, y_1 - y_2) = \overline{\hat{\eta}(x_1,y_1)~\hat{\eta}(x_2,y_2)}$ due to the statistical homogeneity of the Gaussian noise $\hat{\eta}$. Following \cite{Constantinou2015} (see H.4 there), the covariance $Q(x, y)$ is specified on the Fourier domain by the forcing spectrum $\widehat{Q} (m,n) = c(m)^2 d^2 e^{-n^2d^2}$ which is non-zero for zonal wavenumbers $m \in [m_f, m_f + \delta m)$ where $d = 0.1$ and $c(m)$ is such that the net energy injection rate is $\varepsilon$.}

\Cref{fig:sf:qlce2} shows the time-dependent zonal energy obtained when the stochastic forcing described above is applied to two zonal modes $m = [8,9]$ (i.e. $k_f = 8$ and $\delta k = 2$). QL initialised with random noise (\cref{fig:sf:qlce2}a) predicts three unstable zonal modes $m = 3,4$ and $5$ and a stronger mean ($m = 0$, black line). Alongside the forced zonal modes ($m = 8,9$), these are the only significantly energetic zonal modes at $t = 1,000$ days. CE2 initialised with the same IC (\cref{fig:sf:qlce2}b) diverges from the QL solution, predicting a significantly weaker $m = 5$ mode. In contrast to QL, the $m = 6$ mode is among the most energetic modes in CE2. This feature is also  reproduced when CE2 is initialised with the QL end point solution (\cref{fig:sf:qlce2}c). Once again, the complete picture emerges when ranks achieved by second cumulant submatrices are plotted for the diverging solutions. \Cref{fig:sf:qlce2}d compares the ranks of cumulant submatrices for the two CE2 solutions against the unit ranks seen in QL. The forced modes $m = 8,9$ in CE2 take on full ranks as prescribed by the method of forcing, while the remaining modes depart significantly from unity. Once again, we note a departure in the CE2 second cumulant rank from unity in tandem with a divergence between QL and CE2 solutions.

\begin{figure}
    \centering
    \includegraphics[scale=0.9]{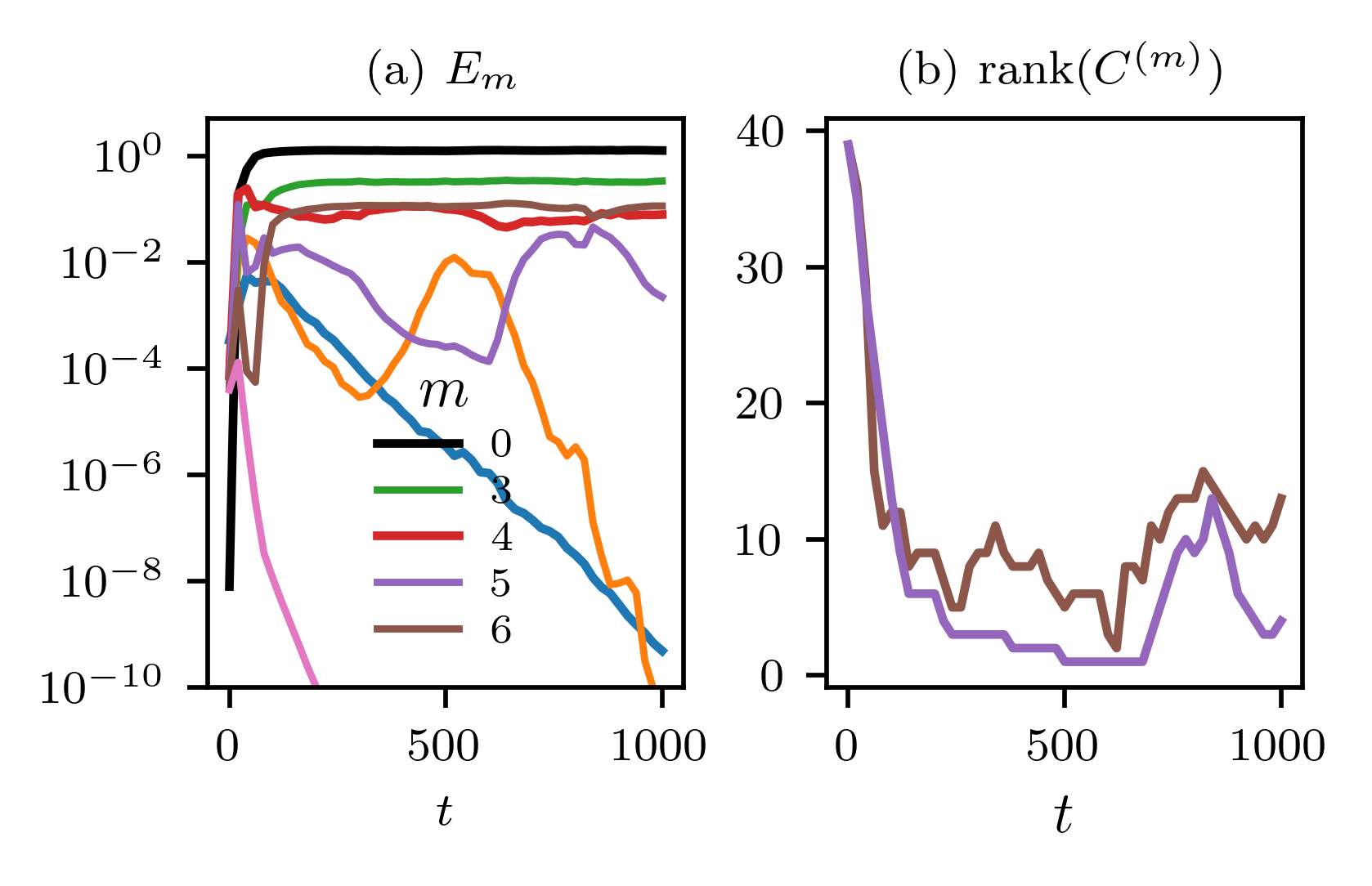}
    \caption{(a) energy in zonal wavenumbers $E_m$ and (b) evolution of the rank of cumulant submatrices $C^{(5)}$ and $C^{(6)}$ with full rank initialisation for the stochastically forced case. Despite the initialisation \emph{and} forcing being full rank, the CE2 solution does not correspond with the QL solution (\cref{fig:sf:qlce2}a).}
    \label{fig:sf:fr}
\end{figure}

We now modify the CE2 solution obtained above (as diverged from QL) to include a full rank (CE2 FR) initialisation of the second cumulant (power $10^{-4}$) coupled with a weakly energetic first cumulant (power $10^{-8}$). The zonal energy evolution in \cref{fig:sf:fr}a indicates that such a CE2 solution returns to the CE2 solutions obtained with unity rank initialisation (\cref{fig:sf:qlce2}b and \cref{fig:sf:qlce2}c), i.e. with a relatively weak $m = 5$ mode (purple) and significant energy in $m = 6$ mode (brown). Hence, CE2 with full rank initialisation also differs from the corresponding QL solution (\cref{fig:sf:qlce2}a). \Cref{fig:sf:fr}b shows that the $m = 5$ mode, though considerably weaker than in QL, initially evolves to a unity rank up until $t = 500$ days -- but the $m = 6$ mode retains a high rank leading it to be considerably more energetic than in QL. These results confirm that the correspondence between the QL state and CE2 statistical dynamics is \emph{not} restored in a system with stochastic (full rank) noise and/or full rank initialisation.

%
\section{Theory}

A CE2 solution can only stay coincident with a QL solution if each second cumulant across the {\em same} excited zonal wavenumbers  stays unit rank. The possibility of a rank instability, which converts a unit rank second cumulant into one which is  multi-rank, is most clearly illustrated with a simple idealisation of the stability problem (for a more detailed discussion see Appendix A of \cite{MK:2023}). Let the linear stability problem for the QL problem at a given zonal wavenumber  be
\begin{equation}
\partial_t {\bf u} = {\bf A} {\bf u}
\label{QL_stab}
\end{equation}
where ${\bf A}$ is a $N \times N$ real matrix and $N$ measures the discretization in $y$. If $\lambda_i$ and ${\bf e}(i)$ are the $ 1\leq i \leq N$ eigenvalues and corresponding right eigenvectors of ${\bf A}$, there will be as many unstable directions as eigenvalues for which $Re(\lambda_i)>0$. The corresponding cumulant perturbation equation for $C_{ij}:=u_i u_j$ is
\begin{equation}
\partial_t {\bf C} = {\bf A} {\bf C}+{\bf C} {\bf A^T}
\label{CE2_stab}
\end{equation}
which has a corresponding matrix eigenvalue problem with $N^2$ eigenvalues. It is straightforward to show that these are made up of all possible (ordered) pairings $\lambda_i+\lambda_j$ with corresponding eigen{\em matrix} ${\bf C}(i,j)={\bf e}(i) {\bf e}(j)$\footnote{Building in the symmetry $C_{ij}=C_{ji}$ as is commonly done reduces the eigenvalue count to $N(N+1)/2$ by not counting $\lambda_i+\lambda_j$ and $\lambda_j+\lambda_i$ separately}.
\NEW{The key observation is if $\lambda_i+\lambda_j$ has a positive real part with $i \neq j$ then there is a new instability direction excitable in CE2 that is {\em not} available to the QL computation \NEW{because} the corresponding eigenvector has no counterpart there\NEW{,} i.e. it is uninterpretable \NEW{in QL}. This is always the case when $\lambda_i$ and $\lambda_j$ have positive real parts but the associated degrees of freedom are already excited separately due to the individual unstable eigenvalues and may not be dynamically important. The situation is different, however, if one unstable eigenvalue partnered with a stable eigenvalue makes the pair unstable in CE2. This is significant because it could lead to stable degrees of freedom in QL being stimulated in CE2. In practice
numerical errors will inevitably excite these new modes -- or rank instabilities -- in CE2 increasing the rank of the cumulant above unity.}

\NEW{A simple example makes this phenomenon clear. Consider the QL system $u_1'=2u_1$ and $u_2'=-u_2$ which has one stable and one unstable direction. The second cumulant is given by
\begin{equation}
C =
\begin{pmatrix}
u_1\\
u_2
\end{pmatrix}
\otimes
\begin{pmatrix}
u_1 & u_2\\
\end{pmatrix}
=
\begin{pmatrix}
u_1^2 & u_1 u_2\\
u_2 u_1 & u_2^2
\end{pmatrix}
\label{rank1}
\end{equation}
and it has one eigenvector with \NEW{non-zero} eigenvalue $u_1^2 + u_2^2$
\begin{equation}
\begin{pmatrix}
u_1\\
u_2
\end{pmatrix}
\end{equation}
and a second eigenvector with zero eigenvalue:
\begin{equation}
\begin{pmatrix}
-u_2\\
u_1
\end{pmatrix}\ .
\end{equation}
The second cumulant \NEW{therefore has a} rank $1$ reflecting the fact that it is constructed from the outer product of a single vector with itself, Eq. \ref{rank1}.}

The equivalent CE2 system where $C_{11} := u_1^2$; $C_{12} = C_{21} := u_1 u_2$ and $C_{22}: = u_2^2$ (and the symmetry $C_{12} = C_{21}$ is built in),
\begin{equation}
\partial_t \left[ \begin{array}{c}
C_{11} \\
C_{12} \\
C_{22}
\end{array}
\right]
=
\left[ \begin{array}{ccr}
4 & 0 & 0 \\
0 & 1 & 0 \\
0 & 0 & -2 \\
\end{array}
\right]
\left[ \begin{array}{c}
C_{11} \\
C_{12} \\
C_{22}
\end{array}
\right]
\end{equation}
has two unstable directions and one stable direction. It is possible to have a perturbation of form $[0\,\, 1\,\, 0]^T$ in the extra unstable direction but this perturbation has no equivalent in the QL problem as it is inconsistent with any one choice of $u_1$ and $u_2$. \NEW{In other words, $C_{12}^2$ need not equal $C_{11}~ C_{22}$ as it must for the QL system (Eq. \ref{rank1}).}

Importantly, a rank instability in CE2 will generally cause the mean flow  to diverge away from its QL equivalent, as seen in the numerical experiments. If this effect is sufficiently strong, it is possible that one system may have different zonal stability properties to the other. In particular, one may be unstable to a new zonal wavenumber whereas the other may not at the same parameter values (e.g. see figure \ref{fig:kf:qlce2}). This zonal instability would further increase the divergence between the two systems.

\NEW{Finally, it is worth remarking that the divergence of QL and CE2 dynamics can occur even if ensemble rather than spatial averaging is used. The only prerequisite is that the quasilinear dynamics are not full ranked (so there are damped degrees of freedom) giving `headroom' for a rank instability in CE2. This is, for example, always the case for a properly resolved QL representation of a dissipative system. To be specific, a QL system consisting of $N$ ODEs which tracks an $n-$dimensional attractor will typically have $n \ll N$  for a well\NEW{-}resolved computation (otherwise $N$ needs to be increased). As a result, an ensemble of all possible initial conditions in ${\mathbb R}^N$ will evolve over time in QL to give a second cumulant of rank $O(n)<N$.  This opens up the possibility of a rank instability occurring in CE2  so that the propagated cumulant has a higher rank than that in the QL computation, thus producing divergence between the systems. Adding stochastic forcing will not change the situation unless the QL dynamics is forced to have a full ranked second cumulant {\em and} all singular values are significant (i.e. the forcing is strong). The latter condition excludes a situation where QL is weakly forced so that the singular values are partitioned into significant values relevant for the inherent dynamical attractor and much weaker values which reflect the passive response to the forcing in otherwise damped degrees of freedom.}\\

%

\section{Conclusion}

Our work indicates that care needs to be taken in interpreting the results of the statistical representation of quasilinear models. The instabilities described above, which can be triggered either by numerical errors, or more physically, by any of the ubiquitous sources of noise such as thermal fluctuations \citep{bandak2021thermal} present in real systems can lead to different solutions emerging for the statistical representation from those that describe any single realisation of a QL flow. Perhaps implicitly recognising this issue, researchers have often initialised CE2 with a full rank second cumulant \citep{pmt2019}  --- often termed a ``maximum ignorance'' initial condition --- and accepted that the resulting statistical description  may not agree with that of any realisation of the underlying QL theory. The instabilities described above indicate that care must be taken in comparing DSS with the results from single realisations of a quasilinear model.  We note that both QL and CE2 are approximations to the full underlying dynamics and it is not {\it a priori} obvious which choice better approximates the statistics of the fully nonlinear system. Higher-order truncations of cumulant expansions (such as CE2.5 and CE3) lead naturally to a higher-rank second cumulant, owing to the feedback of the higher cumulants on the evolution of the second cumulant \citep[see e.g. Ref.][]{Marston2019}. We conclude by stressing that these results are in no way a criticism of either Direct Statistical Simulation at CE2 or investigations utilising a single realisation of a quasilinear model. We simply emphasise that these types of investigation can only elucidate different processes. Realisations of a QL model are useful for determining which nonlinear processes may be important in the dynamics of a turbulent flow, whilst statistical models such as CE2 yield information about the expected (in a statistical sense) response of system.

\section*{Acknowledgements}

We acknowledge support of funding from the European Union Horizon 2020 research and innovation programme (grant agreement no. D5S-DLV-786780).  JBM and SMT are supported in part by a grant from the Simons Foundation (Grant No. 662962, GF).

\bibliographystyle{jfm}
\bibliography{references}

\end{document}